\documentclass[twocolumn,prb,aps,showpacs,floatfix,amssymb,nobibnotes]{revtex4}
\usepackage{amsmath}%
\usepackage{bm}%
\usepackage{graphicx}
\usepackage{color}

\begin{document}

\title{Unique magnetic structure of YbCo$_2$Si$_2$}

\author{N. Mufti$^{1,2}$, K. Kaneko$^{1,3}$, A. Hoser$^4$, M. Gutmann$^5$, C. Geibel$^1$, 
C. Krellner$^{1,6}$, and O. Stockert$^1$}

\affiliation{$^1$Max-Planck-Institut f\"ur Chemische Physik fester Stoffe, 01187 Dresden, Germany}
\affiliation{$^2$Department of Physics, State University of Malang, Malang 65145, Indonesia}
\affiliation{$^3$Quantum Beam Science Center, Japan Atomic Energy Agency, Tokai, Naka, Ibaraki 319-1195, Japan}
\affiliation{$^4$Helmholtz-Zentrum Berlin f\"ur Materialien und Energie, 14109 Berlin, Germany} 
\affiliation{$^5$ISIS Neutron and Muon Source, Rutherford Appleton Laboratory, Harwell Oxford, Didcot OX11 0QX, United Kingdom}
\affiliation{$^6$Physikalisches Institut, Goethe-Universit\"at Frankfurt, 60438 Frankfurt, Germany}

\date{\today}

\begin{abstract}
We report on the results of powder and single crystal neutron diffraction to investigate the magnetic order in YbCo$_2$Si$_2$ below the N\'eel temperature $T_{\rm N} = 1.7$\,K in detail. Two different magnetically ordered phases can clearly be distinguished. At lowest temperatures a commensurate magnetic structure with a propagation vector ${\bf k}_1 = (0.25~ 0.25~ 1)$ is found, while the intermediate phase ($T > 0.9$\,K) is characterized by an incommensurate magnetic structure with ${\bf k}_2 = (0.25~ 0.086~ 1)$. The magnetic structure in YbCo$_2$Si$_2$ is in marked contrast to all other known RCo$_2$Si$_2$ compounds (R = rare earth element) likely due to some itineracy of the Yb 4f states being responsible for the magnetism. 
\end{abstract}
\pacs{71.27.+a, 75.25.-j, 61.05.F-}

\maketitle

\section{Introduction}
Yb-based compounds have attracted particular interest due to their unique physical properties such as intermediate-valence, heavy-fermion, and recently non-Fermi-liquid (NFL) phenomena in the vicinity of quantum critical points (QCPs) \cite{temmerman99,custers03,gegenwart08}. One of the most intensively studied Yb-based heavy-fermion compounds is YbRh$_2$Si$_2$ which orders antiferromagnetically at a very low $T_{\rm N} \approx $ 0.07 K \cite{trovarelli00}. The magnetic order can be suppressed to zero by application of a small critical magnetic field  or by applying slightly negative chemical pressure through substitution of La, Ir, or Ge  for Yb, Rh and Si, respectively \cite{custers03,gegenwart08,trovarelli00,ferstl05,friedemann09}. Resistivity and thermodynamic measurements of this compound show interesting properties such as pronounced NFL behavior \cite{trovarelli00}, divergence of the heavy quasiparticle mass and of the magnetic Gr\"uneisen parameter at the QCP \cite{custers03,tokiwa09}. However, the very weak antiferromagnetic order in YbRh$_2$Si$_2$ makes it difficult to study the unusual properties using microscopic probes. Several attempts to determine the magnetic structure by neutron diffraction failed due to the small size of the ordered magnetic moment (10$^{-3}$ $\mu_B$) \cite{ishida03}. A further complication in performing neutron scattering on YbRh$_2$Si$_2$ is the high neutron absorption cross section of Rh restricting the effective sample thickness. Nevertheless, dynamic magnetic correlations were detected at incommensurate positions $(\delta~ \delta~0)$, $\delta \approx 0.14$ at low temperatures $T = 0.3$\,K exhibiting no dispersion \cite{stock12}. These incommensurate spin correlations evolve into ferromagnetic correlations at a higher temperature of $15$\,K \cite{stock12}. 

To overcome the difficulties in studying pure YbRh$_2$Si$_2$ we followed the route to stabilize the antiferromagnetic state by application of hydrostatic pressure or by suitable chemical substitution. The latter reduces the unit cell volume and acts as positive chemical pressure. This results in an increase of both, the ordering temperature $T_{\rm N}$ and the size of the ordered moment. Our studies indeed show that substitution of Rh by isovalent Co stabilizes the antiferromagnetic state and enhances $T_{\rm N}$ \cite{klingner11,westerkamp08}. Recently, we reported neutron powder diffraction results on YbCo$_2$Si$_2$ \cite{kaneko10} where we observed for the first time magnetic superstructure peaks at lowest temperature which exhibit a pronounced shift in an intermediate phase below $T_{\rm N}$. The intensities of these two phases show significant changes and point to marked differences in the magnetic structures \cite{kaneko10}. However, in those initial experiments we could not determine unambiguously the propagation vector, since only very few magnetic peaks could be observed. Therefore, a more detailed understanding of the magnetic ordering phenomena in YbCo$_2$Si$_2$ is highly desired.
Furthermore, the magnetic structure in YbCo$_2$Si$_2$ serves as a reference for iso-structural quantum-critical YbRh$_2$Si$_2$.

The magnetic properties of YbCo$_2$Si$_2$ (space group $I4/mmm$) were initially studied on polycrystalline samples by Hodges who observed antiferromagnetic order below T$_{\rm N}$ $\approx $ 1.7 K with an ordered magnetic moment of $1.4\,\mu_{\rm B}$/Yb using M\"{o}ssbauer spectroscopy measurements \cite{hodges87}. He interpreted the reduced moment in comparison to e.g. the free Yb$^{3+}$ moment as a result of crystalline electric field (CEF) effects. Later on, magnetization \cite{kolenda89,klingner11a} and inelastic neutron powder measurements \cite{goremychkin00} confirmed the Yb valence to be trivalent and revealed a CEF level scheme of this tetragonal compound with a ground state doublet and excited doublets at $\Delta \approx 4.0$, $12.5$ and $30.5$\,meV. Recently, we studied single-crystalline YbCo$_2$Si$_2$ using magnetic susceptibility, magnetization, specific heat and electrical resistivity measurements \cite{mufti10,pedrero10,klingner11,klingner11a}. As a result, YbCo$_2$Si$_2$ displays a pronounced easy-plane anisotropy and exhibits two magnetic transitions at $T_{\rm N} \approx  1.7$\,K and $T_{\rm L} \approx 0.9$\,K. The high-temperature susceptibility follows a Curie-Weiss behavior with an effective moment $\mu_{eff} \approx 4.7\,\mu_{\rm B}$/Yb in close agreement with the free Yb$^{3+}$ moment and a Weiss temperature $\Theta \approx -4$ and $160$\,K for magnetic fields $B$ parallel and perpendicular to the basal plane, respectively. No magnetic contribution from Co has been observed \cite{klingner11,kolenda89,klingner11a}. Magnetization, susceptibility and magnetoresistance measurements indicate a complex $B-T$ phase diagram with some anisotropy in the basal plane \cite{pedrero10,mufti10}. Here we report the results of our investigation from powder and single crystal diffraction measurements to study the magnetic structure of the two magnetically ordered phases in YbCo$_2$Si$_2$.

\section{Experiment}
Single crystals and polycrystalline samples of YbCo$_2$Si$_2$ were prepared by flux growth using indium as flux \cite{klingner11}. X-ray powder diffraction was used to verify the crystal structure. 
Neutron powder diffraction experiments were performed at the Berlin Neutron Scattering Center (BENSC) using the diffractometer E2 with a neutron wavelength  $\lambda =  2.39$\,{\AA}. Approximately $m = 7$\,g of YbCo$_2$Si$_2$ powder was sealed inside a copper container. A small amount of a deuterated methanol-ethanol mixture in the sample container, which solidifies upon cooling, improved the thermal coupling, but also resulted in an increased background and a broad hump around $2\Theta = 40$\,{deg.~}as seen for all pattern in Fig.\,\ref{figure1}. The sample can was mounted on a $^3$He stick and allowed measurements in the temperature range between $T = 0.4$ and $2$\,K. 
Typical counting times per powder pattern displayed in Fig.\,\ref{figure1} were $7.5$ hours.
The powder pattern were analyzed by the Rietveld method using FullProf \cite{rodriguez-carvajal93}.
Single-crystal neutron diffraction was carried out on the instrument SXD at the ISIS facility which utilizes the time-of-flight Laue technique \cite{keen06}. A YbCo$_2$Si$_2$ single crystal of approximate dimensions $5\times4\times0.5$\,mm$^3$ ($m \approx 50$\,mg) was mounted on copper foil and attached to the cold plate of a $^3$He cryostat. Data were collected at temperatures $T = 0.45$, $1.1$  and $2$\,K.
In addition, an elastic neutron scattering experiment was performed on 
the triple-axis spectrometer TAS-2, installed at the research reactor JRR-3 in Tokai, Japan.
Neutrons with a wavelength $\lambda =  2.359$\,{\AA} ($E = 14.7$\,meV) were used together with a horizontal collimation open-80'-80'-open and a pyrolithic graphite (PG) filter on the incident beam was used to eliminate higher-order contamination. The single crystal with a mass of $m \approx 38$\,mg was mounted on a copper post connected to the cold finger of a $^3$He cryostat with the $[001]$ axis vertical resulting in a $(h~ k~Ê0)$ horizontal scattering plane. Data were taken on TAS-2 at temperatures $T = 0.27 - 2$\,K with typical counting times of $\approx 30$\,s per point. All neutron intensities are normalized to a monitor ("mon") in the incident neutron beam.

\section{Results and discussion}

The x-ray powder diffraction pattern of YbCo$_2$Si$_2$ obtained at room temperature as well as the neutron powder diffraction pattern taken at $2$\,K confirm the tetragonal ThCr$_2$Si$_2$-type structure ($I4/mmm$) in which the Yb atoms occupy the 2a site, while the Co atoms sit on the 4d site and Si atoms are placed on the 4e site with $z= 0.3650(1)$. The refined lattice parameters from x-ray diffraction at room temperature are $a =  3.8512(1)$\,{\AA} and $c = 9.6887(2)$\,{\AA} in good agreement with the previously reported values by Kolenda et al.\cite{kolenda89}.

\begin{figure}[t]%
\centering
\includegraphics*[width=\linewidth]{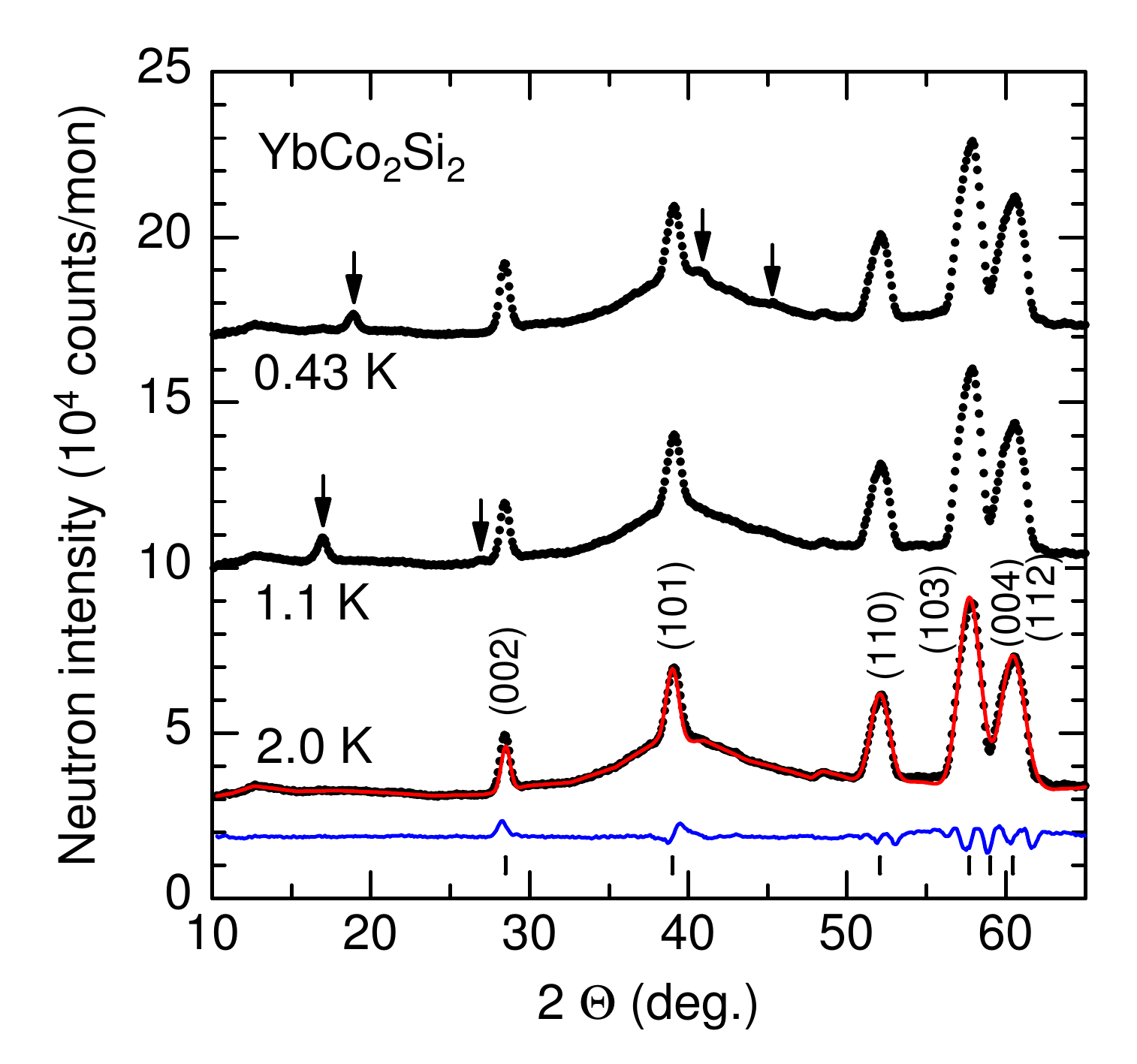}
\caption{Powder neutron diffraction pattern of YbCo$_2$Si$_2$ recorded at $T = 0.43$, $1.1$ and $2$\,K on E2 with a neutron wavelength $\lambda = 2.39$\,{\AA}. The nuclear peaks are labeled and the ticks at the bottom denote their position. The arrows indicate magnetic superstructure peaks. The fit to the $2$\,K data is shown by a solid red line and the deviation of the fit from the data is displayed by the blue line. Data for different temperatures are shifted vertically with respect to each other.}
\label{figure1}
\end{figure}

\begin{figure*}[t]
\includegraphics[width=0.95\linewidth,clip]{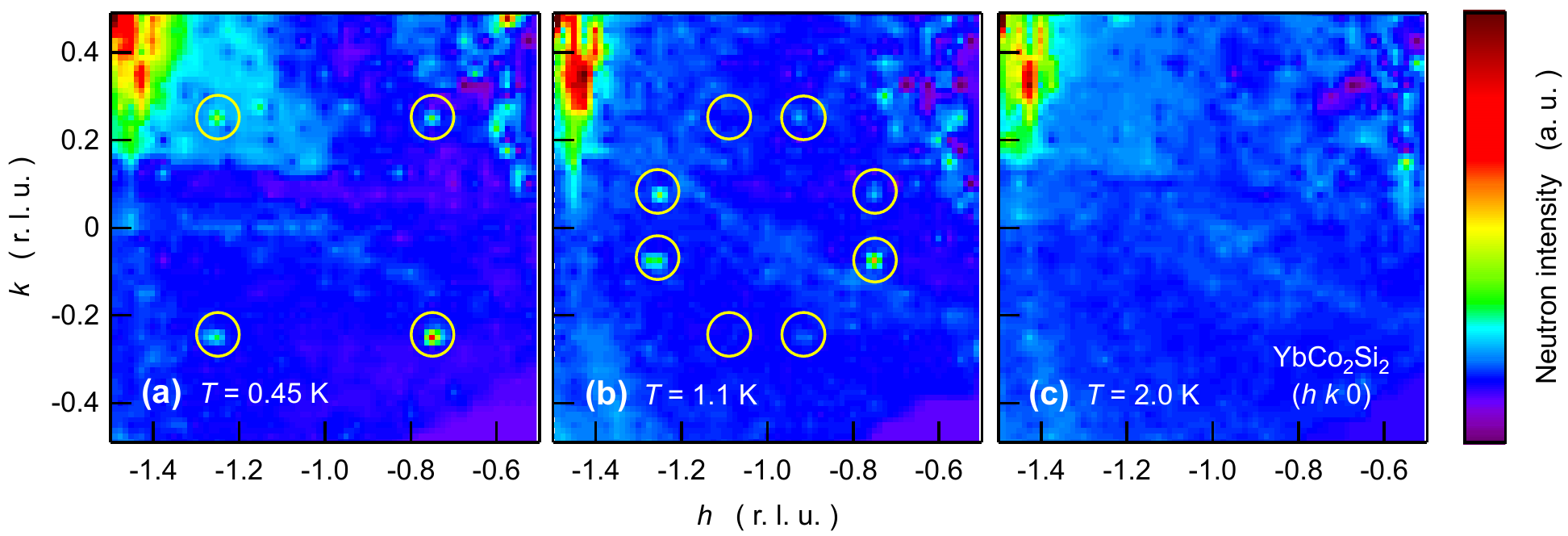}
\caption{Color-coded intensity map of the reciprocal $(hk0)$ plane of YbCo$_2$Si$_2$ (a) at $T = 0.45$\,K in the ground state, (b) for $T = 1.1$\,K in the intermediate phase and (c) in the paramagnetic regime at $T = 2$\,K. Additional magnetic peaks are visible in (a) and (b) and are highlighted by yellow circles. Likely due to domain population the intensities at the magnetic $(-1.08~\pm0.25~0)$ positions in (b) are quite low. The intensity maps have been extracted from the raw single-crystal neutron diffraction data recorded on SXD.}
\label{figure2}
\end{figure*}

Fig.\,\ref{figure1} displays the neutron powder pattern of YbCo$_2$Si$_2$ taken at three different temperatures corresponding to the paramagnetic regime at $T = 2$\,K and in the two magnetically ordered phases at $T = 0.43$ and $1.1$\,K. The measurements are in line with the previous neutron diffraction results \cite{kaneko10} and reveal additional Bragg peaks below $2$\,K due to magnetic ordering. However, only three magnetic peaks at lowest temperature and two magnetic peaks in the intermediate phase are clearly visible making it difficult to unambiguously determine the propagation vector. For a reliable determination of the propagation vector single-crystal neutron diffraction is inevitable and was carried out on SXD. 
 The neutron intensity in a part of the reciprocal $(hk0)$ scattering plane in YbCo$_2$Si$_2$ is shown as color-coded intensity maps in Fig.\,\ref{figure2} for $T = 0.45$, $1.1$ and $2$\,K. Magnetic superstructure peaks have been observed in the ground state at $T = 0.45$\,K and in the intermediate phase at $T = 1.1$\,K and are marked by yellow circles. They can be indexed as $(-1.25~\pm0.25~0)$ and $(-0.75~\pm0.25~0)$ in the ground state whereas the magnetic peaks in the intermediate phase are located at $(-1.25~\pm0.08~0)$, $(-0.75~\pm0.08~0)$, $(-1.08~\pm0.25~0)$ and $(-0.92~\pm0.25~0)$ (cf. Fig.\,\ref{figure2} (a) and (b)). The intensities at $(-1.08~\pm0.25~0)$ and $(-0.92~-0.25~0)$ are very low, much weaker than at the other symmetry-equivalent positions, likely due to domain population.

\begin{figure}[t]
\includegraphics[width=0.95\linewidth,clip]{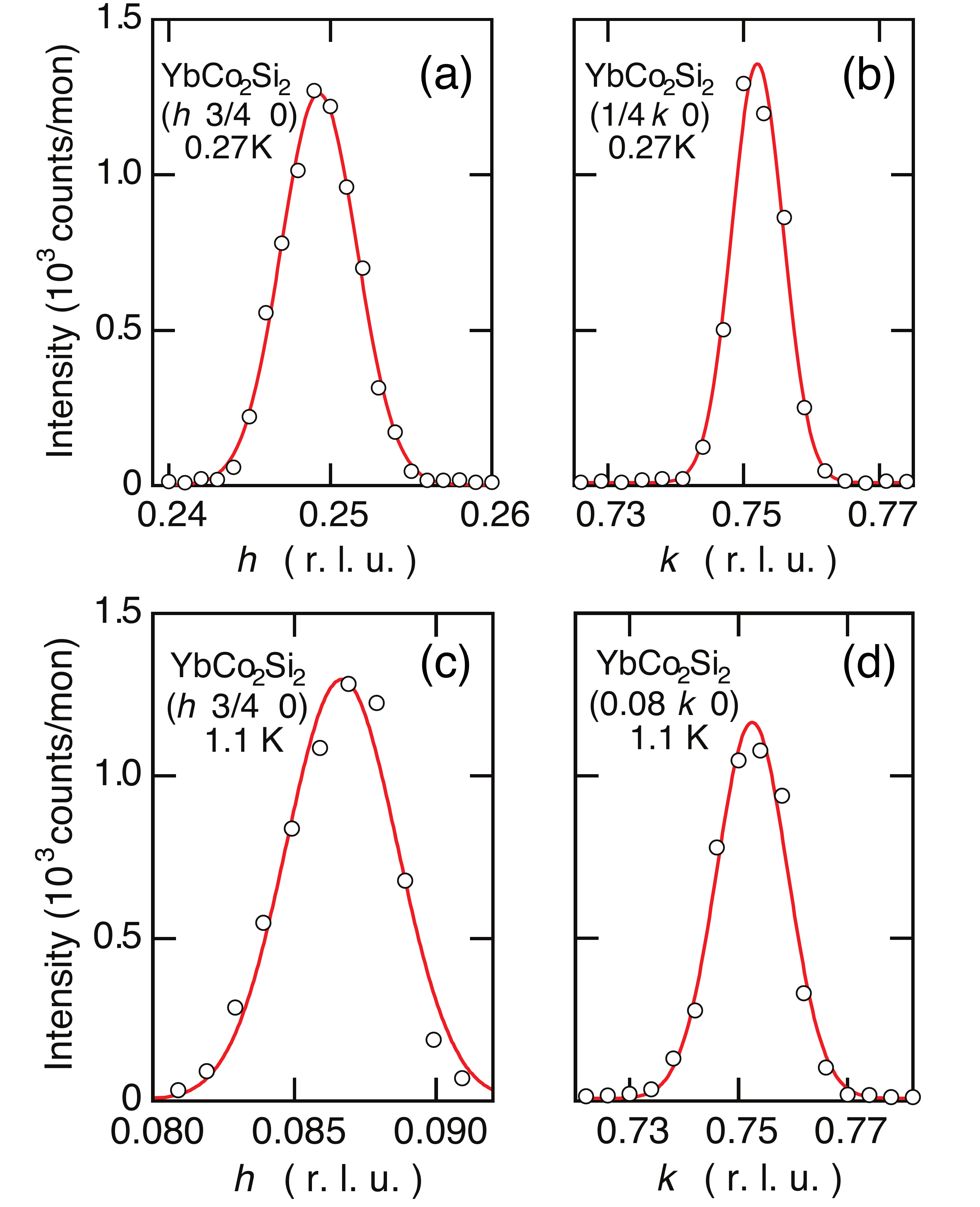}
\caption{Scans along $[100]$ and $[010]$ across (a,b) the magnetic peak at ${\bf Q} \approx (0.25~ 0.75~ 0)$ at $T = 0.27$\,K and (c,d) the magnetic reflection at ${\bf Q} \approx (0.086~ 0.75~Ê0)$ in the intermediate phase at $T = 1.1$\,K. Solid lines indicate fits of the magnetic peaks with Gaussian lineshape. Data taken on TAS-2.}
\label{figure3}
\end{figure}

The neutron data taken on TAS-2 allowed a more accurate determination of the propagation vector. Fig.\,\ref{figure3} shows scans along $[100]$ and $[010]$ across magnetic peaks in reciprocal space in the ground state and in the intermediate phase. At lowest temperature $T = 0.27$\,K one magnetic peak was observed at the commensurate position $(0.25~0.75~Ê0)$ (cf. Fig.\ref{figure3}(a) and (b)). Fig.\,\ref{figure3} (c) and (d) reveal the position of a magnetic peak in the intermediate phase to be at $(0.086~0.75~Ê0)$. Taking the positions of several magnetic peaks together, these single-crystal measurements give clear evidence that the magnetic order at lowest temperatures is commensurate with a propagation vector ${\bf k}_1 = (0.25~ 0.25~ 1)$ while the intermediate phase is characterized by an incommensurate structure with ${\bf k}_2 = (0.25~ 0.086~ 1)$. Equivalent to ${\bf k}_2 = (0.25~ 0.086~ 1)$ is the propagation vector $(0.086~Ê0.25~Ê1)$ when interchanging the $a^*$- and $b^*$-component of ${\bf k}_2$ within the tetragonal structure. No change of the incommensurate value of ${\bf k}_2$ within the measurement accuracy has been detected between $T_L$ and $T_{\rm N}$.
 The absence of $(100)$ and $(001)$ magnetic reflections indicates that the magnetic structure in both phases is not as simple as for other members of the family of RCo$_2$Si$_2$ compounds (R= Pr, Nd, Ho, Tb and Dy) whose magnetic order can be represented as an antiferromagnetic stacking of ferromagnetic $ab$-planes along the tetragonal $c$-axis \cite{leciejewicz83}. 
A reliable determination of magnetic intensities from the single-crystal measurements was not possible because absorption cannot be neglected for the very thin, plate-like crystals, and due to
the effect of domain formation and population in this highly symmetrical, tetragonal compound.

\begin{figure}[t]%
\centering
\includegraphics[width=\linewidth,clip]{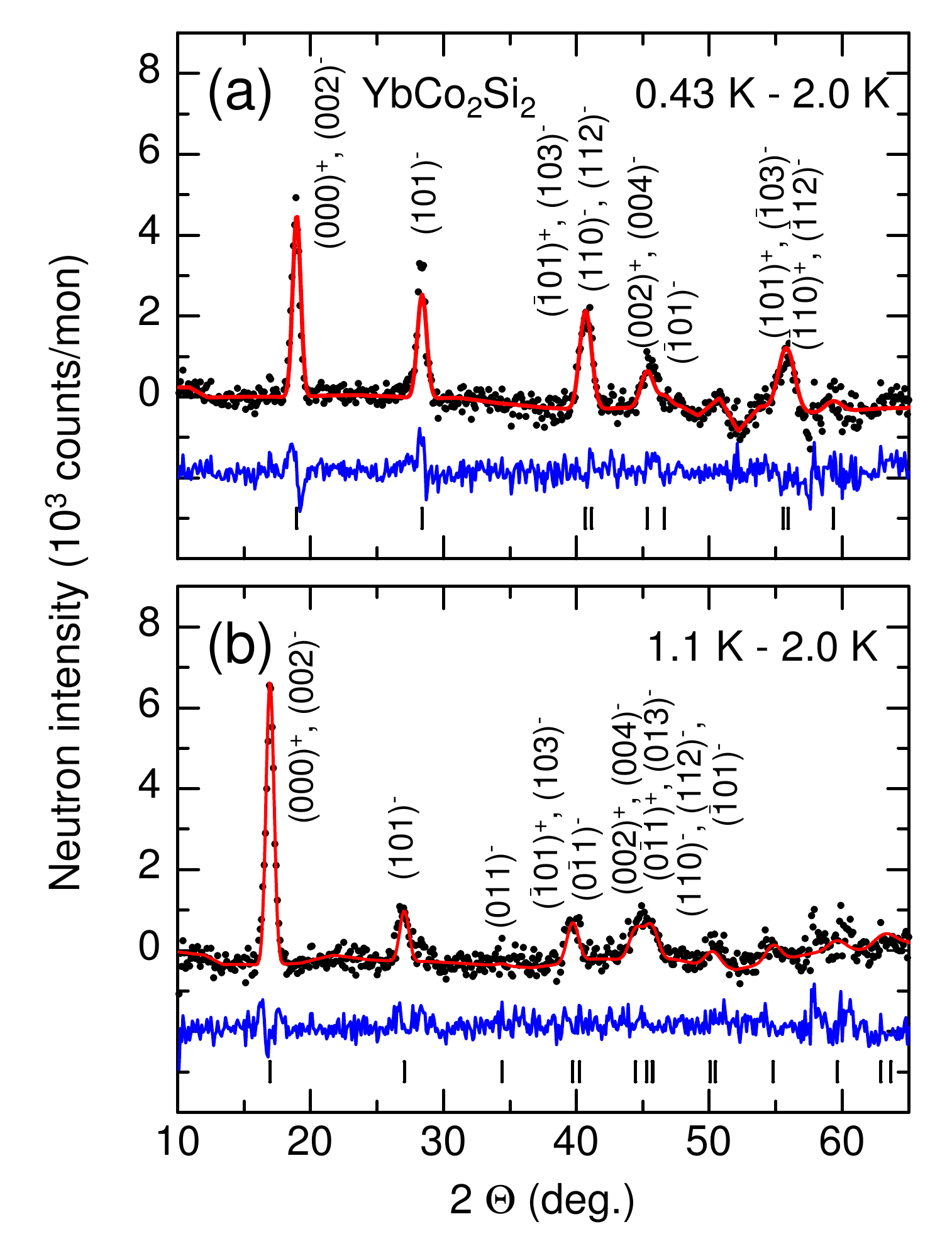}
\caption{Difference neutron diffraction pattern of YbCo$_2$Si$_2$ (a) in the ground state at $T= 0.45$\,K and (b) in the intermediate phase at $T = 1.1$\,K obtained by subtracting the powder pattern measured at $T = 2.0$\,K (data of Fig.\,\ref{figure1}). Red solid lines indicate fits of the magnetic structure to the data (see main text) while blue lines show the deviations of the fits from the experimental data. The row of ticks at the bottom of each panel denote the magnetic peaks corresponding to the appropriate propagation vectors ${\bf k}_1 = (0.25~ 0.25~ 1)$ in (a) and ${\bf k}_2 = (0.25~ 0.086~ 1)$ in (b).}
\label{figure4}
\end{figure}

Instead, the neutron powder pattern of YbCo$_2$Si$_2$ were reinvestigated with the knowledge of the propagation vectors. In order to separate the magnetic contribution to the powder pattern, the data recorded in the paramagnetic regime at $T = 2$\,K were subtracted from the pattern at lower temperatures. The resulting difference pattern are displayed in Fig.\,\ref{figure4} together with fits of the magnetic structure to the data (see below). The propagation vectors are validated by the powder measurements and the positions of the observed magnetic satellite peaks are well described by ${\bf k}_1$ and ${\bf k}_2$. Moreover, the existence of magnetic satellite peaks of the $(000)$ and $(002)$ nuclear peaks in both phases indicates that a considerable fraction of the ordered moment is perpendicular to the c-axis. This is in good agreement with previous results from the M\"{o}ssbauer spectroscopy \cite{hodges87}. 

\begin{figure}[t]%
\centering
\includegraphics*[width=0.95\linewidth,clip]{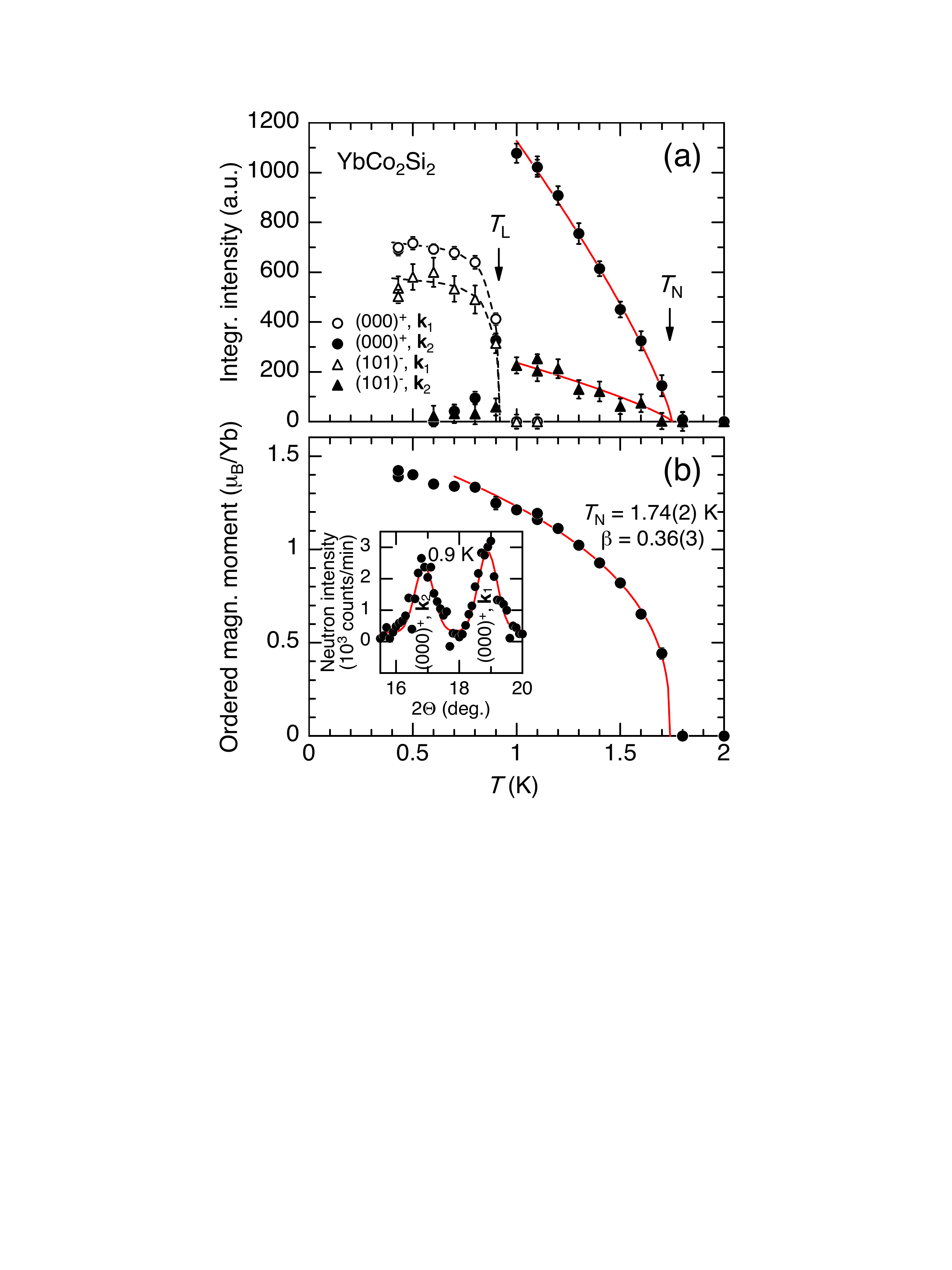}
\caption{%
(a) Integrated intensity of several magnetic superstructure peaks versus temperature in YbCo$_2$Si$_2$ as deduced from the powder diffraction data. (b) Temperature dependence of the ordered magnetic moment as extracted from fits to the whole powder diffraction pattern. Dashed lines are guides to the eyes, while the red solid lines are power-law fits to the data as described in the main text. Inset: Low-angle part of difference neutron diffraction pattern at $T = 0.9$\,K together with fit (solid line) showing coexisting superstructure peaks of both magnetic phases.}
\label{figure5}
\end{figure}

The two strongest magnetic superstructure peaks in the powder diffraction pattern of YbCo$_2$Si$_2$, the $(000)^+$ and $(101)^-$ satellite peaks [$(hkl)^\pm = (hkl) \pm {\bf k}_{1/2}$ denotes a satellite peak of the nuclear $(hkl)$ peak], have been recorded at several temperatures. Their integrated intensity as a function of temperature is plotted in Fig.\,\ref{figure5}(a). Magnetic intensity fully vanishes just above $1.7$\,K in close agreement with bulk measurements \cite{klingner11}. 
In contrast to the continuous phase transition at $T_{\rm N}$, the intensity of the commensurate magnetic peaks disappears suddenly at $T_L \approx 0.9$\,K. Moreover, as displayed in the inset of Fig.\,\ref{figure5}(b) the pattern recorded at $T = 0.9$\,K shows the magnetic superstructure peaks of both phases, the commensurate as well as the incommensurate one, i.e., phase separation takes place indicating a first-order transition from the commensurate low-temperature to the incommensurate intermediate-temperature phase. Our results are consistent with the first-order nature of this transition seen in heat capacity and magnetization measurements \cite{klingner11,pedrero11}.


\begin{figure}[t]%
\centering
\includegraphics*[width=\linewidth]{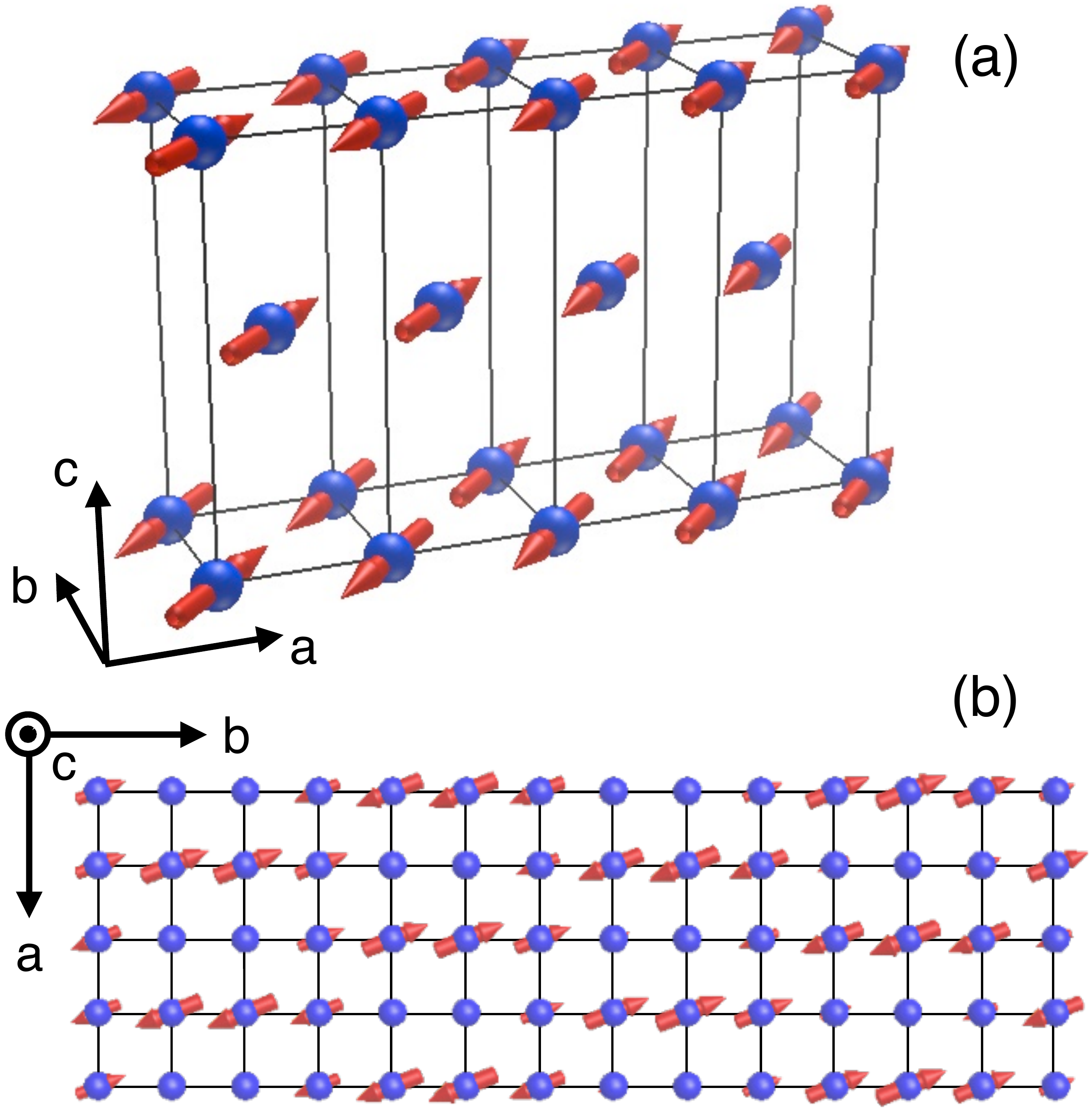}
\caption{%
     Magnetic structure of the Yb sublattice in YbCo$_2$Si$_2$ (a) in the ground state at $T < T_{L} \approx 0.9$\,K and (b) the proposed amplitude-modulated magnetic structure (only tetragonal basal plane displayed) in the intermediate, incommensurate phase at $T_{L} < T < T_{\rm N} \approx 1.7$\,K with moments in the basal plane.}
\label{figure6}
\end{figure}

We now turn to the discussion of the magnetic structure in YbCo$_2$Si$_2$ in more detail. Magnetic neutron scattering only probes magnetic moment components perpendicular to the scattering vector. Hence, the strong increase of the $(000)^+$ magnetic intensity from the low-temperature to the intermediate-temperature phase is mainly due to a spin reorientation taking place at the commensurate -- incommensurate transition. The change in the propagation vector from ${\bf k}_1$ to ${\bf k}_2$ seems only to be a minor effect.

In order to determine the magnetic structure a magnetic symmetry analysis has been performed to find the allowed irreducible representations and their basis vectors for the propagation vectors ${\bf k}_1$ and ${\bf k}_2$ in the space groups $I4/mmm$ (for details see Appendix). Such an analysis for the YbCo$_2$Si$_2$ magnetic structure (Yb$^{3+}$ on 2a site of the tetragonal $I4/mmm$ unit cell and a magnetic propagation vector ${\bf k}_1 = (0.25~ 0.25~ 1)$) yields as possible directions for the magnetic moments (the basis vectors of the irreducible representations) the $[110]$, $[1 \overline{1} 0]$ or the $[001]$ direction (cf. Table \ref{table1}). Moments along $[001]$ can be excluded, since the $(002)^+$ magnetic satellite peak would then be almost vanished. However, this magnetic peak at $2\Theta \approx 46$\,deg. is clearly visible. Furthermore, magnetization measurements indicate that YbCo$_2$Si$_2$ is an easy-plane antiferromagnet \cite{pedrero11}. To decide whether the magnetic moments are perpendicular or (anti-)parallel to the projection of the propagation vector ${\bf k}_1$ on to the basal plane, we fitted the powder pattern for both possible moment directions. Assuming moments along $[1 \overline{1} 0]$ we could only get the intensity of the primary $(000)^+$ peak correct while the intensities of all magnetic satellites at higher scattering angles are systematically strongly underestimated. In contrast, assuming magnetic moments in $[110]$ direction a satisfactory fit to all magnetic peaks can be obtained as seen in Fig.\,\ref{figure4}(a). With ${\bf k}_1 = (0.25~ 0.25~ 1)$ and moments along $[110]$ we obtain a commensurate antiferromagnetic structure with equal moments on each Yb site. The size of the ordered moment at $T = 0.43$\,K is determined to $1.42(2)\,\mu_{\rm B}$/Yb in close agreement with the M\"ossbauer results. Fig.\,\ref{figure6}(a) is a visualization of the magnetic structure where only the modulation of the magnetic moments along $a$ is shown. Along the $b$ direction the same modulation of the moments occurs as along $a$. 

Above $T_{\rm L} \approx 0.9$\,K, i.e., in the incommensurate intermediate phase with ${\bf k}_2  = (0.25~ 0.086~ 1)$, moments along $[001]$ can be ruled out for similar reasons as in the low-temperature phase. However, symmetry analysis allows also for magnetic moments in the tetragonal basal plane along $[100]$ and/or $[010]$. Considering an amplitude-modulated magnetic structure and the moment direction along $[100]$ the calculation yields for the two most prominent magnetic satellites at lowest scattering angles (cf. Fig.\,\ref{figure4}(b)) a ratio of the intensities $I_{(000)^+}/I_{(101)^-} \approx 360$ while the ratio of the measured intensities amounts to $\approx 7.4$. Hence, such a model with magnetic moments only along $[100]$ can also be discarded. Assuming that the Yb moments are amplitude modulated and point in the $[010]$ direction results in a ratio $I_{(000)^+}/I_{(101)^-} \approx 6.8$ being quite close to the observation. The fit of this model to the powder pattern is displayed in Fig.\,\ref{figure4}(b) and describes already well the data. It yields for the ordered magnetic moment $1.19(1)\,\mu_{\rm B}$/Yb (root mean square value) at $1.1$\,K. It should be noted that no higher harmonic magnetic satellite peaks, i.e., {\bf G}$\pm$3{\bf k$_2$}, {\bf G}$\pm$5{\bf k$_2$}, ... (${\bf G}$ being a reciprocal lattice vector) have been observed. 
Allowing the magnetic moments to take any direction within the tetragonal basal plane as permitted by symmetry (see Appendix), either an amplitude-modulated magnetic structure or an in-plane spiral spin arrangement fit equally well the experimental data. Only with the help of polarized neutron diffraction can both structures be distinguished and the incommensurate magnetic structure unambiguously be determined. 
The fit of the amplitude-modulated structure gives a moment in the basal plane forming an angle of about $20$\,deg. with the $[010]$ axis.
The modulation of the magnetic Yb$^{3+}$ moments within the basal plane of YbCo$_2$Si$_2$ is displayed in Fig.\,\ref{figure6}(b). 
As a result, the magnetic moments in the intermediate phase are aligned almost perpendicular to the propagation vector ${\bf k}_2  = (0.25~ 0.086~ 1)$ and hence to $(000)^+$ which is the origin of the strong increase in intensity of the $(000)^+$ peak when crossing the commensurate -- incommensurate phase boundary.
Since thermodynamic and transport properties display strong anisotropies within the basal plane with quite different magnetic $B-T$ phase diagrams \cite{mufti10,pedrero11}, we favor the amplitude-modulated magnetic structure instead of a spiral magnetic structure for the intermediate, incommensurate phase. However, as already mentioned above polarized neutron diffraction is desired to fully solve the structure in the incommensurate phase. The
intensities of the few magnetic peaks recorded in the single crystal measurement are consistent with the magnetic structure derived from the neutron powder measurements. 

The temperature dependence of the ordered magnetic moment has been extracted from fits of the two above-mentioned magnetic structures to the whole powder pattern taken at different temperatures and is shown in Fig.\,\ref{figure5}(b). The ordered moment increases smoothly below $T_{\rm N}$ without any major anomaly at $T_L$ corroborating that at $T_L$ just a spin reorientation takes place.
A power-law fit of the form $m=m_0(1-(T/T_{\rm N}))^{\beta}$ to the ordered magnetic moment (solid line in Fig.\,\ref{figure5}(b)) describes well the data from about $1$\,K up to $T_{\rm N}$. It
yields a N\'eel temperature $T_{\rm N} = 1.74(2)$\,K and a critical exponent $\beta = 0.36(3)$, the latter being quite close to $\beta = 0.367$ expected for a 3D Heisenberg system, but still compatible with $\beta = 0.345$ within the 3D $X\!-\!Y$ model \cite{collins89}.
As expected, with the same $T_{\rm N}$ and $\beta$ the temperature dependence of the $(000)^+$ and $(101)^-$ peaks ($I=I_0(1-(T/T_{\rm N}))^{2\beta}$), whose integrated intensities were obtained independently by single peak fits, is reproduced as indicated by the solid lines in Fig.\,\ref{figure5}(a). It is shown above that the magnetic moments of the antiferromagnetic structure are confined to the basal plane as expected from the CEF analysis and the anisotropy in the magnetic susceptibility \cite{klingner11a}. Hence a 3D $X\!-\!Y$ model might be appropriate for YbCo$_2$Si$_2$. However, the almost isotropic critical magnetic fields to suppress antiferromagnetic order \cite{pedrero10,mufti10} suggest a Heisenberg scenario. Further measurements are needed to decide which model is better suited to describe YbCo$_2$Si$_2$.

Comparing the magnetic structure of YbCo$_2$Si$_2$ with other members of the RCo$_2$Si$_2$ (R = rare earth Pr, Nd, Tb, Dy, Ho, Er and Tm) series similarities, but also marked differences become apparent. Well below their individual ordering temperatures all rare-earth based RCo$_2$Si$_2$ except YbCo$_2$Si$_2$ order antiferromagnetically with a propagation vector ${\bf k} = (0~ 0~ 1)$ in a collinear magnetic structure. Ferromagnetically ordered basal planes are stacked antiferromagnetically along the $c$-direction and break the body centering of the tetragonal crystal structure \cite{leciejewicz83,leciejewicz83a}. While the magnetic moments in the light and the heavy rare-earth based RCo$_2$Si$_2$ up to HoCo$_2$Si$_2$ are aligned along $c$, from ErCo$_2$Si$_2$ on and including YbCo$_2$Si$_2$ the moments are oriented perpendicular to the $c$-axis. This can be understood as a result of the different CEF schemes as determined by inelastic neutron scattering and M\"{o}ssbauer spectroscopy \cite{goerlich82,leciejewicz83,leciejewicz83a,goremychkin00,harker02}. 
The origin lies in the sign of the $B_{20}$ parameter (in Stevens notation) in the CEF Hamiltonian, which is related to the paramagnetic Curie-Weiss temperatures along the different crystallographic directions.
Depending on the sign of the $B_{20}$ parameter a change from a large uniaxial magnetic anisotropy (magnetic moments along the $c$-direction) to a basal-plane anisotropy (moments perpendicular to the $c$-axis) occurs. Common to all RCo$_2$Si$_2$ is the existence of an incommensurate modulated magnetic phase just below the N\'eel temperature $T_{\rm N}$ \cite{shigeoka88,shigeoka89,szytula00,schobinger03} whose origin lies in the RKKY-interaction, the oscillatory exchange interaction between the rare-earth 4f electrons mediated by the conduction electrons. 

The interplay between the RKKY-interaction, favoring incommensurate modulated magnetic structures, and the CEF effects, preferring certain moment directions, determines the experimentally realized magnetic structures in the RCo$_2$Si$_2$ compounds. In contrast to all other RCo$_2$Si$_2$, YbCo$_2$Si$_2$ exhibits unusual magnetic propagation vectors. In YbCo$_2$Si$_2$ not only an antiferromagnetic stacking of the Yb$^{3+}$ moments occurs along $c$, but also the ordering in the basal plane is antiferromagnetic. Moreover, both propagation vectors, ${\bf k}_1 = (0.25~ 0.25~ 1)$ and ${\bf k}_2  = (0.25~ 0.086~ 1)$, together with moments in the basal plane break the tetragonal symmetry of YbCo$_2$Si$_2$. It remains currently an open question to which extent the symmetry of the crystallographic structure is also reduced in the magnetically ordered regime.
The occurrence of two quite different magnetically ordered phases in YbCo$_2$Si$_2$ indicates that both magnetic structures are energetically close to each other. This suggests that the magnetic interactions in YbCo$_2$Si$_2$ might be frustrated. Especially for the quantum-critical sister compound YbRh$_2$Si$_2$ the possible existence of frustration is an important issue and will be further pursued in the future.

While the magnetic order in the other RCo$_2$Si$_2$ is of purely localized moments, the unusual magnetic behavior of YbCo$_2$Si$_2$ might well be a result of the fact that the 4f electrons of Yb are not fully localized, but retain some small itinerant character. A small amount of itinerant Yb 4f states has indeed been recently seen in ARPES measurements on YbCo$_2$Si$_2$ \cite{guettler14} and is also corroborated by heat capacity and resistivity measurements \cite{klingner11}.
Interestingly, a similar effect is observed in the isovalent Kondo lattice CeRh$_2$Si$_2$. This compound presents a complex antiferromagnetic order of seemingly completely localized Ce$^{3+}$ moments \cite{grier84,kawarazaki00}, but with propagation vectors in marked difference to those of the RRh$_2$Si$_2$ (R = rare earth) homologues with a stable rare earth local moment. In CeRh$_2$Si$_2$ the width of the quasi-elastic line in neutron scattering in the paramagnetic state indicates a significant Kondo interaction \cite{severing89a}. However, de Haas--van Alphen experiments did not reveal a strong effect of this Kondo interaction in the electronic states at the Fermi level \cite{araki01}. Thus, in the two compounds YbCo$_2$Si$_2$ and CeRh$_2$Si$_2$, where the RKKY interaction succeeds over the Kondo effect, the latter does not lead to strong changes in the electronic states observed directly, but seems to strongly affect the propagation vector of the magnetically ordered state. Therefore, the propagation vector might be a very sensitive probe of hybridization effects in Kondo lattices with dominating RKKY interaction.

\section{Conclusion}

We have presented results of single crystal and powder neutron diffraction on YbCo$_2$Si$_2$ to study the magnetic order below $T_{\rm N} \approx 1.7$\,K in detail. The magnetic structure in the ground state is a collinear structure with a propagation vector ${\bf k}_1 = (0.25~ 0.25~ 1)$ and ordered magnetic moments of $1.4\,\mu_{\rm B}$/Yb oriented along $[110]$. In the intermediate phase above $T_{\rm L} \approx 0.9$\,K the magnetic order becomes incommensurate with ${\bf k}_2  = (0.25~ 0.086~ 1)$ and either a spiral spin arrangement in the basal plane or amplitude-modulated moments in the basal plane almost along $[010]$. Magnetic intensity vanishes continuously at the N\'eel temperature $T_{\rm N} \approx 1.7$\,K in accordance to thermodynamic and transport measurements and exhibits a critical exponent $\beta = 0.36(3)$ compatible with a 3D $X\!-\!Y$ or Heisenberg system. The exceptional propagation vectors in YbCo$_2$Si$_2$ within the RCo$_2$Si$_2$ series might be related to the partial itinerant character of the Yb 4f states.

We greatly acknowledge fruitful discussions with M. Baenitz, M. Brando, A. Hannaske, S. Kirchner, and L. Pedrero. We thank C. Klausnitzer and R. Weise for technical support. Access to the ISIS spallation neutron source and the Berlin BER-II and Tokai JRR-3 reactor neutron sources is gratefully acknowledged. This work was financially supported by the DFG Research Unit 960 ``Quantum Phase Transitions'' and a Grant-in-Aid for Scientific Research (C) (No. 2454036) from the Japan Society of the Promotion of Science.

\section*{Appendix}

\begin{table}[b]
\begin{tabular}{| c | c p{3ex} c|}
  \hline
  Propagation & Irreducible &  & Basis\\
  Vector {\bf k} & Representation & & Vector \\
  \hline
 & $\Gamma_2$ & &$\Psi_1 = [1 1 0]~$ \\
 \cline{2-4}
$(0.25~ 0.25~ 1)$ &$\Gamma_3$ & &$\Psi_2 = [1 \bar{1} 0]~$ \\
 \cline{2-4}
 & $\Gamma_4$ & &$\Psi_3 = [0 0 1]~$ \\
  \hline
  \hline
 &  $\Gamma_1$ & &$\Psi_1 = [0 0 1]~$ \\
 \cline{2-4}
$~(0.25~ 0.086~ 1)~$ &$\Gamma_2$ & &$\Psi_2 = [1 0 0]~$ \\
 &  $\Gamma_2$ & &$\Psi_3 = [0 1 0]~$ \\
  \hline
\end{tabular}
\caption{Irreducible representations and associated basis vectors for Yb$^{3+}$ on the 2a site of the $I4/mmm$ structure in YbCo$_2$Si$_2$ with a propagation vector ${\bf k}_1 = (0.25~0.25~1)$ or ${\bf k}_2 = (0.25~0.086~1)$.}
\label{table1}
\end{table}

\begin{figure}[th]%
\centering
\includegraphics[width=\linewidth,clip]{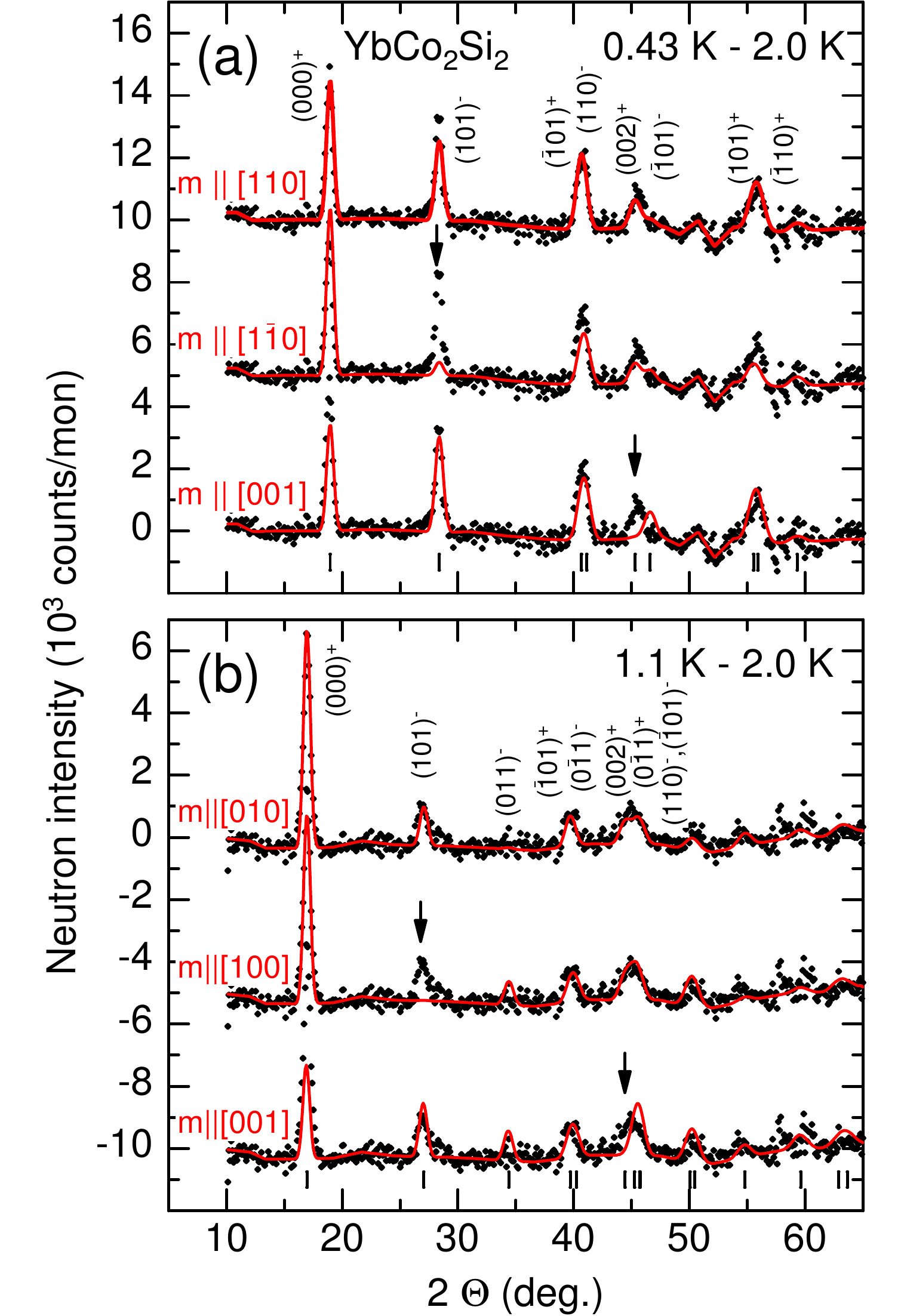}
\caption{Difference neutron diffraction pattern of YbCo$_2$Si$_2$ (a) in the ground state at $T= 0.45$\,K and (b) in the intermediate phase at $T = 1.1$\,K obtained by subtracting the powder pattern measured at $T = 2.0$\,K (data of Fig.\,\ref{figure1}). Solid lines indicate fits of different magnetic structure models to the data. Arrows indicate some pronounced deviations of the fits from the data.  The row of ticks at the bottom of each panel denote the magnetic peaks corresponding to the appropriate propagation vectors ${\bf k}_1 = (0.25~ 0.25~ 1)$ in (a) and ${\bf k}_2 = (0.25~ 0.086~ 1)$ in (b). In addition, strong magnetic peaks are indexed. Different model fits (together with the corresponding data) are shifted vertically with respect to each other.}
\label{figure7}
\end{figure}

For the determination of the magnetic structure we performed a representation analysis for both ground state and intermediate phase using the program BasIreps being part of the FullProf suite \cite{rodriguez-carvajal93}. As a consequence of the Landau theory for a second-order phase transition such an analysis implies that only one irreducible representation becomes critical. Then, for the magnetic structure determination only the moment directions given by the basis vectors which belong to the irreducible representation being considered, have to be taken into account. Since the transition at the N\'eel temperature is second order, representation analysis certainly is valid for the intermediate phase ($T_L < T < T_{\rm N}$). However, even for the commensurate magnetic structure at lowest temperatures representation analysis holds since this state can also be reached from the paramagnetic state at high fields at low $T$ via a second-order transition as a function of (decreasing) magnetic field along $[110]$ \cite{pedrero11}.

The representation analysis for the magnetic Yb on the 2a site within the tetragonal $I4/mmm$ structure of YbCo$_2$Si$_2$ and a propagation vector ${\bf k}_1 = (0.25~ 0.25~ 1)$ yields for the magnetic representation three one-dimensional irreducible representations $\Gamma_{Mag}  = \Gamma_2 + \Gamma_3 + \Gamma_4$ following Kovalev's notation \cite{kovalev93}. The basis vectors $\Psi$ of these irreducible representations indicate possible directions of the ordered magnetic moment. Table\,\ref{table1} displays the irreducible representations as well as the basis vectors.  

For the intermediate phase ($T_L < T < T_{\rm N}$) the magnetic representation decomposes into two one-dimensional representations, one of which occurs twice, $\Gamma_{Mag}  = \Gamma_1 + 2\Gamma_2$. The corresponding basis vectors together with their irreducible representations are again shown in Table\,\ref{table1}. In addition to an amplitude-modulated structure equally possible is also a spiral structure in the intermediate phase since for the $\Gamma_2$ representation moments along $[100]$ and/or $[010]$ are allowed (see main text).

The different allowed magnetic structures have been fitted to the data, i.e., magnetic moments along $[110]$, $[1 \overline{1} 0]$ or $[001]$ as given by $\Gamma_2$, $\Gamma_3$ or $\Gamma_4$ for the commensurate magnetic structure and moments along $[001]$ or in the basal plane ($[100]$ and/or $[010]$) for the incommensurate intermediate phase ($\Gamma_1$ or 2 $\Gamma_2$ representations). Fits are displayed in Fig.\,\ref{figure7} together with the magnetic diffraction pattern already shown in Fig.\,\ref{figure4}. Clearly the commensurate structure is described by magnetic moments along $[110]$ ($\Gamma_2$ representation), i.e. parallel to the basal-plane components of the propagation vector, while the intermediate phase is characterized by moments in the basal plane ($\Gamma_2$ representation) with moments (almost) along $[010]$, i.e. (almost) perpendicular to the propagation vector.

%

\end{document}